# Investigating Current State-of-The-Art Applications of Supportive Technologies for Individuals with ADHD

Fatemah Husain

## 1. Introduction

It is considered normal for most students to attend classes on time, for most travelers to plan for a trip without problems, and for most movie watchers to watch a movie completely at once. However, for 3.4% of adults (Fayyad et al., 2007) and for 7.2% of children (Thomas et al., 2015) it is nearly impossible to be on time for classes or other important meetings; arranging travel details is also considered a complex task; and staying engaged while sitting still watching a 2-hour movie is almost impossible as well. Those people who find those tasks very difficult to complete represent individuals with Attention Deficit Hyperactivity Disorder (ADHD). For them to accomplish these tasks, it takes more effort, and often requires alternative approaches that better suit their mental processing model. Having diverse ways of thinking and processing information is not to be criticized, but to be embraced within every aspect of human-centered design systems (Gernsbacher, 2007).

Individuals from various ages might get diagnosed with ADHD through three core symptoms of inattention, hyperactivity, and impulsivity. Those symptoms are reflected in activities performed at various settings, including the workplace, home, and school. To help in reducing the effects of ADHD symptoms, there are multiple treatments, but none of them help in curing ADHD. The four most common treatments for ADHD include medications (e.g., Ritalin, Adderall, and Dexedrine), behavioral interventions (e.g., verbal reinforcement, peer mediation, and visual cues), skills training (e.g., coaching, role-playing, and social skills group) and behavioral therapies (e.g., parent-child interaction therapy, parent management training, and cognitive behavioral therapy). Technological solutions offer great opportunities in delivering treatments, especially those related to behavioral interventions, monitoring, and modification in a more flexible, acceptable, and accessible way (e.g., wearable devices, smartphone applications, video games, and social robotics). Moreover, using technology to support the treatments has the benefit of enabling comparison of behavioral changes overtime, especially when the technology supports recording user data and visualization features.

Early support during childhood prevents the exacerbation of behavioral disorders before entering adulthood. Comparing the percentages of research on different mental disabilities for children, illuminated the lack of research on ADHD. The percentage of studies involving children with Autism Spectrum Disorder (ASD) is 64.1%, while those that involve children with ADHD is 3.3% (Börjesson et al., 2015). However, the percentage of children with ADHD is increasing; the estimated number of children with ADHD in the USA was 5.4 million in 2007, 6.4 million in 2011, and 6.1 million in 2016 (National survey of children's health, 2017, September 6). Few researchers focus on children with ADHD, especially those working on Human-Computer Interaction (HCI). The HCI researchers study the design of computer systems, their accessibility to diverse populations, and the interaction of the users with the systems. Users are a key component in the design of assistive technologies, thus for the HCI researchers to tackle the needs of individuals with ADHD it is important that they fully understand user needs in order to develop supportive systems that can interface, communicate, and interact with users effectively and efficiently by considering their mental models.

Reviewing previous work on assistive technologies for individuals with ADHD offers valuable opportunities to researchers on assistive technology by providing detailed investigation of current state of the art applications of research covering the same topic, pointing out existing problems, and showing gaps in previous research, which could be used to support and advance the research on and development of ADHD technological solutions.

The purpose of this study is to perform a Systematic Literature Review (SLR) to elicit the state of the art on assistive technologies for individuals with ADHD within HCI research domain by focusing on the following research questions:

1. What are the challenges and ages of individuals with ADHD included in the design process? and in which contexts are they included?
2. What methods and techniques are typically used when including individuals with ADHD in the design process, and what are the sample sizes of participants?
3. What types of technologies are mostly developed, and who are the targeted populations?
4. What design frameworks and protocols for involving individuals with ADHD in the design have been described?

In addressing the questions above, an SLR was conducted following the Kitchenham model for computing literature from the Association Computing Machinery's Digital Library (ACM DL). Literatures from the HCI domain (e.g., system design and interface) were included, while literatures from other domains were excluded, such as algorithms or computer vision. Publications between January 2007 and December 2017 were evaluated.

## 2. DEFINING THE TARGET POPULATION AND DESIGN METHODS

### 2.1 Individuals affected by ADHD

ADHD stands for Attention Deficit Hyperactivity Disorder, it is considered a chronic mental and behavioral disorder that interferes with everyday activities. ADHD is a "neurodevelopmental disorder defined by persistent impairing levels of inattention, motor hyperactivity and impulsivity that exhibit a negative impact in functioning" (Durães et al., 2015, P.:1861). The main symptoms of ADHD are inattention, hyperactivity, and impulsivity. The American Psychiatric Association (APA) defines three types of ADHD: inattentive, hyperactive/impulsive, or combined type. ADHD is more common among children and its symptoms decrease as they get older (Musser et al., 2016).

Individuals with ADHD face multiple difficulties. Staying focused and paying attention to details is very hard for them, which causes learning disabilities and difficulties in completing tasks. Executive Functioning (EF) skills, such as goal oriented tasks, organization and time management, are of the most challenging and result in difficulty being on time for meetings, submitting assignments, performing exams, etc. The difficulties they face everyday make them very stressed resulting in sleep disorders on most cases. Table 1 describes ADHD challenges based on studies done by the National Resource on ADHD.

*Table 1 ADHD challenges and their prevalence (National Resource on ADHD)*

| | Challenges | Prevalence |
|---|---|---|
| **Neurodevelopmental Disorder** | Learning Disorder | 1 out of 2 |
| | Tourette Syndrome | 1 out of 10 |
| | Speech Problem | 1 out of 10 |
| **Co-occurrence Condition** | Anxiety | 1 out of 2 |
| | Sleep Problems | 1 out of 5 |
| | Substance Abuse | 1 out of 10 |
| **Behavioral Disorder** | Conduct Disorder | 1 out of 4 |
| | Oppositional Defiant disorder | 1 out of 2 |
| **Mood Disorder** | Depression | 1 out of 10 |
| | Bipolar Disorder | 1 out of 5 |

Studies included in this analysis are the ones that target people affected by ADHD. Individuals with ADHD, either adults or children, are the most dominant population, followed by parents and caregivers who are responsible for taking care of individuals with ADHD. Teachers are also considered within the targeted population since they are responsible for children with ADHD while they are at school, they need assistive tools to adapt to their learning difficulties.

**2.2 Design Methods**

Researchers conducting studies on ADHD adopt multiple methods in performing their studies, such as interviewing, surveying, prototyping, or brainstorming, which might include experts and individuals with ADHD. These actions represent the design methods of the study which is used in this analysis. Since this SLR focuses on studies involving the interaction and design of systems, most of the design methods adopted are Participatory Design (PD) and User-Centered Design (UCD) in which individuals with ADHD are the main participants.

In this study, the analysis depends on the classification of design methods defined by Börjesson et al (2015) based on the three main system design phases: requirements, design, and evaluation. A combination of multiple methods can be used within the same study. Furthermore, since some researchers build upon findings from previous studies rather than collecting the requirements personally, which is not present in Börjesson et al (2015) classification model, findings from previous research was added to the requirements phase methods. Table 2 describes the resulting categorization of design methods used in this study.

*Table 2 Classification of methods and techniques (based on Börjesson et al., 2015)*

| Requirements | Design | Evaluation |
|---|---|---|
| Survey | Brainstorming | Participatory or assisted evaluation |
| Field study/ User observation | Creative Sessions | User testing |
| Diary keeping | Card sorting | Questionnaires |
| User requirement interview | Prototyping (Paper, software, wizard-of-Oz) | Post-experience interviews |
| Focus group | | |
| Previous study findings | | |

## 3. RESEARCH METHOD

A Systematic Literature Review (SLR) was conducted following the Kitchenham model (Kitchenham, 2004; Kitchenham et al., 2007; Brereton et al, 2007; Kitchenham et al., 2013). Figure 1 shows the main phases of an SLR along with their activities. Following an SLR model in conducting research has multiple benefits, including its iterative process and defining clear protocols which make it easier for other researchers to redo the study, or depend on its procedures and findings to extend it to a new study. Figure 1 describes the Kitchenham SLR three main phases: planning the review, conducting the review, and reporting the findings.

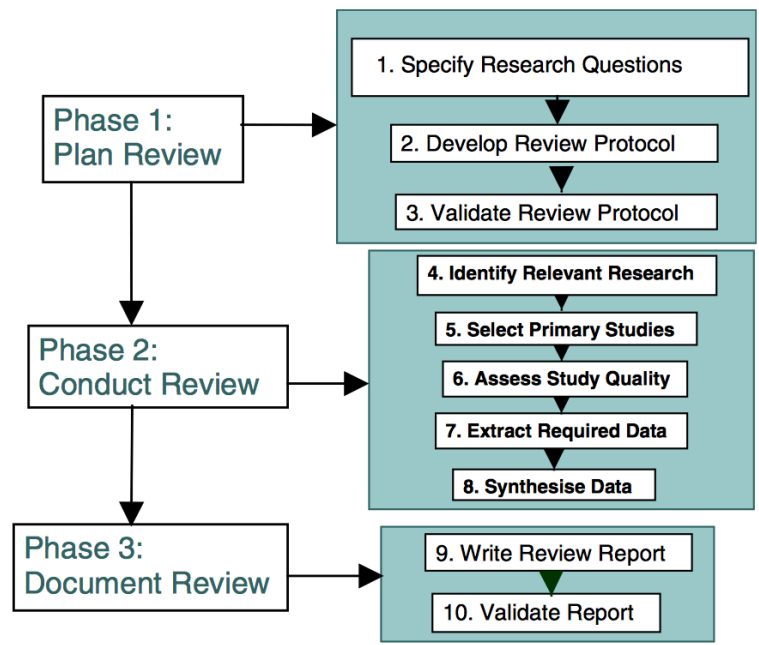

*Figure 1 The Kitchenham's systematic literature review process model (Brereton et al., 2007, P. 572)*

## 3.1 Planning Review

### 3.1.1 Aims and Objectives

The main objective of this literature review is to investigate state-of-the-art of assistive technology involving individuals with ADHD, and from the HCI domain. The scope of the study is limited to literature had published in the Association for Computing Machinery Digital Library (ACM DL).

### 3.1.2 Research Questions

To better understand the state of the current literature covering ADHD, the following four questions were used to guide this analysis:

1. What are the challenges and ages of individuals with ADHD included in the design process? and in which contexts are they included?
2. What methods and techniques are typically used when including individuals with ADHD in the design process, and what are the sample sizes of participants?
3. What types of technologies are mostly developed, and who are the targeted populations?
4. What design framework and protocols for involving individuals with ADHD in the design have been described?

## 3.2 Conducting the Review

### 3.2.1 Strings/Keywords Selection

Three keywords were selected, or strings as named by the Kitchenham model, to conduct this study. The keywords used include "ADHD" which must be present anywhere in text. In addition, two more optional keywords "technology" and "game" were used in which at least one of them must be present anywhere in text.

### 3.2.2 Inclusion and Exclusion Criteria

English papers that are accessible online and published in peer-reviewed journals between January 2007 and December 2017 were included in this review. Papers that are not focusing on the design or interaction of artifacts and technologies with the target group, and dissertations and theses were excluded. Table 3 shows the excluded literatures and the reason of exclusion.

*Table 3 Excluded literature*

| No. | Excluded Paper | Reason |
|---|---|---|
| 1 | Adeli, 2017 | Not related to system design & interaction – Data Mining |
| 2 | Ansari, 2010 | Not related to system design & interaction – Computer Vision |
| 3 | Anthony et al., 2012 | Workshop paper |
| 4 | Anuradha et al., 2010 | Not related to system design & interaction – Algorithm |
| 5 | Barricelli & Loreto, 2017 | Workshop paper and not related to ADHD |
| 6 | Chu et al., 2016 | Not related to system design & interaction – Machine Learning |
| 7 | Cmiel et al., 2011 | Not related to system design & interaction – Biomedical Engineering |
| 8 | Hu et al., 2016 | Not related to system design & interaction – Algorithm |
| 9 | Kushlev et al., 2016 | Not related to ADHD |
| 10 | Lee & An, 2011 | Not related to system design & interaction – Algorithm |
| 11 | Morris et al., 2015 | Not related to system design & interaction – Technical Workforce |
| 12 | Murphy, 2005 | Before 2007 |
| 13 | Rincon et al., 2017 | Not related to system design & interaction – Management |
| 14 | Russo, 2018 | Abstract only |
| 15 | Spicker et al., 2016 | Not related to ADHD |
| 16 | Szykman et al., 2015 | Not related to ADHD |
| 17 | Vilor-Tejedor et al., 2016 | Not related to system design & interaction – Algorithm |

3.2.3 Data Analysis and Extraction

Before Begin with the review process, tables were created with the main attributes to be extracted as a template for data extraction step. Papers were coded into themes to be categorized based on main addressed problem. Exclusion and inclusion criteria were applied after reading the entire paper. From each paper, the following data were extracted: publication details (e.g., authors, publication year), goal of the study, name and description of the technological solution developed, details about the participants age, challenges addressed, methods and techniques used, target group, sample size, results, limitation, and design guidelines.

## 3.3 Documenting the Results

### 3.3.1 Trends on Research of Individuals with ADHD

For the period between January 2007 and December 2017, the most active year with the maximum number of studies on technologies for individuals with ADHD was 2016 in which 8 papers were found. The total number of literature satisfies inclusion criteria was 27, 17 literatures were excluded. Most of the excluded literatures were not related to the HCI, they are not discussing issues on designing systems, accessibility methods, or interacting with systems. Figure 2 shows the number of literature per year which shows that researchers start investigating this domain of science more significantly from 2013.

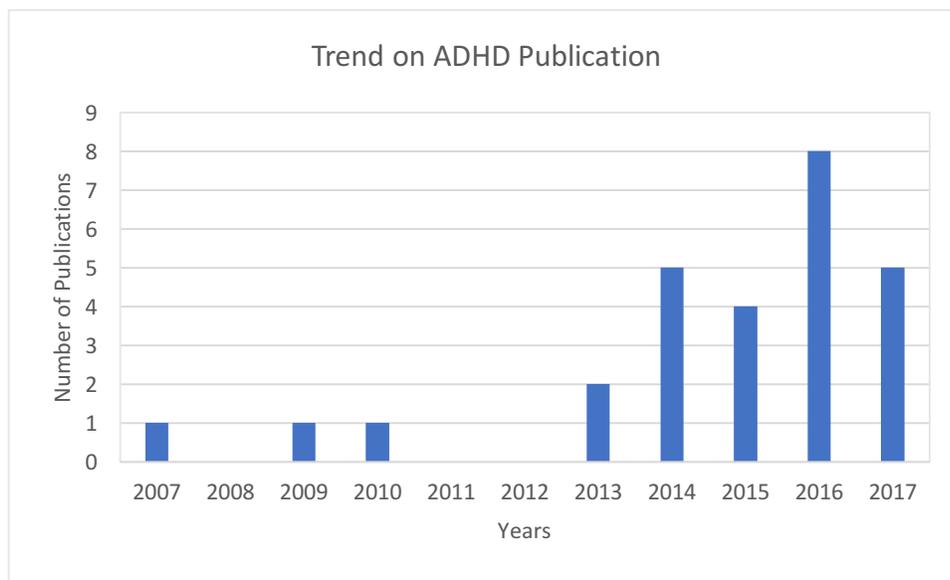

*Figure 2 Number of literature per year related to system design and human interaction with computers for individuals with ADHD*

3.3.2 Challenges/Themes and Target Populations

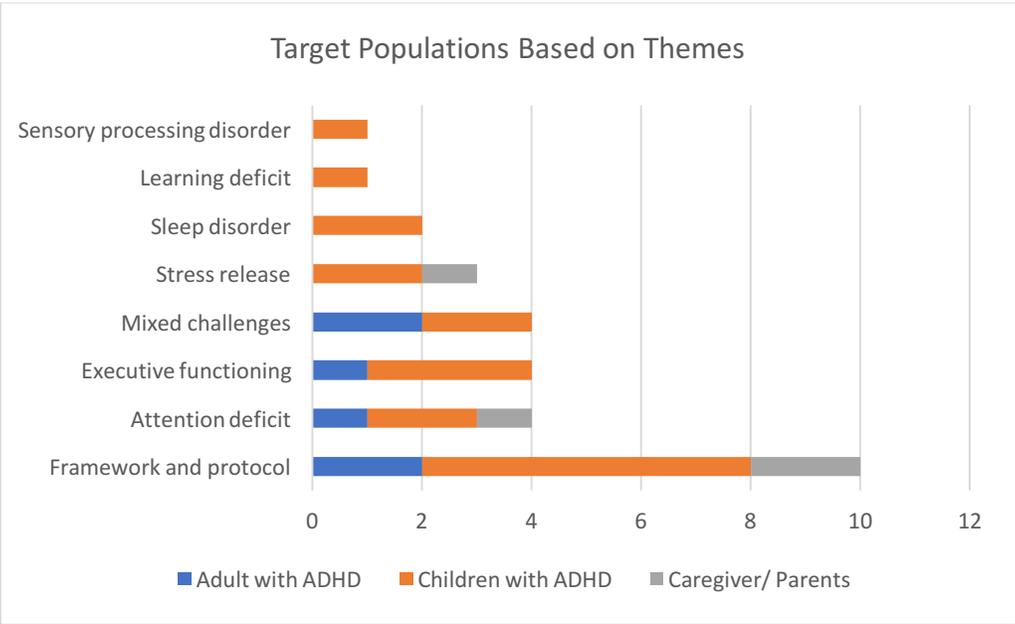

*Figure 3 Number of papers based on themes and target populations*

After reading each paper, the main theme of the contents was defined. Then, papers were categorized based on their main theme, which in most cases addresses the challenges of individuals with ADHD. Based on the papers included that satisfy the inclusion criteria, six themes were identified to categorize all papers: attention deficit, Executive Functioning (EF) skills which affects individuals' abilities in starting new task, stopping it, or switching among multiple tasks (Huh & Ackerman, 2010), framework and protocol including design guidelines, learning deficit, mixed challenges including papers that address more than one challenges and difficulties, Sensory Processing Disorder (SPD) which describes the case where individuals with ADHD become unable to interpret sensory signals correctly (Duvall et al., 2016), sleep disorder, and stress release.

Three main groups for population were identified based on findings from literature to represent the end users: adults with ADHD, children with ADHD, and caregivers or parents that are taking care of the individuals with ADHD. None of the included papers cover the needs of teachers of children with ADHD. Each group of users have different needs and goals which require specific solutions and methods in studying. Figure 3 and Table 4 describe the number of studies covered each theme based on target populations. It is observable that most of the studies for all populations are discussing design frameworks and protocols. Moreover, children were the dominant targeted population, whereas studies for caregivers and parents were limited even though ADHD is hereditary (Pina et al., 2014; Vilor-Tejedor et al., 2016), on approximately 75% of the cases (Faraone, 2014), which means that many parents are also having some problems which make it harder to support their children in the treatments.

Among the included studied, 13 studies involve children below 18 years old within the design or evaluation methods. These studies follow the participatory design approach, and the maximum number of studies per age (in a yearly basis) was 5 studies for children between 9 and 10 years old. None of these studies had participants under 5 years old. That limitation in involving young children in the design approach could be related to difficulties in diagnosing ADHD symptoms during or before this age, or to difficulties in explaining the procedures of the study while accomplishing the design methods. Figure 4 shows the distribution of the number of studies based on the children yearly age group.

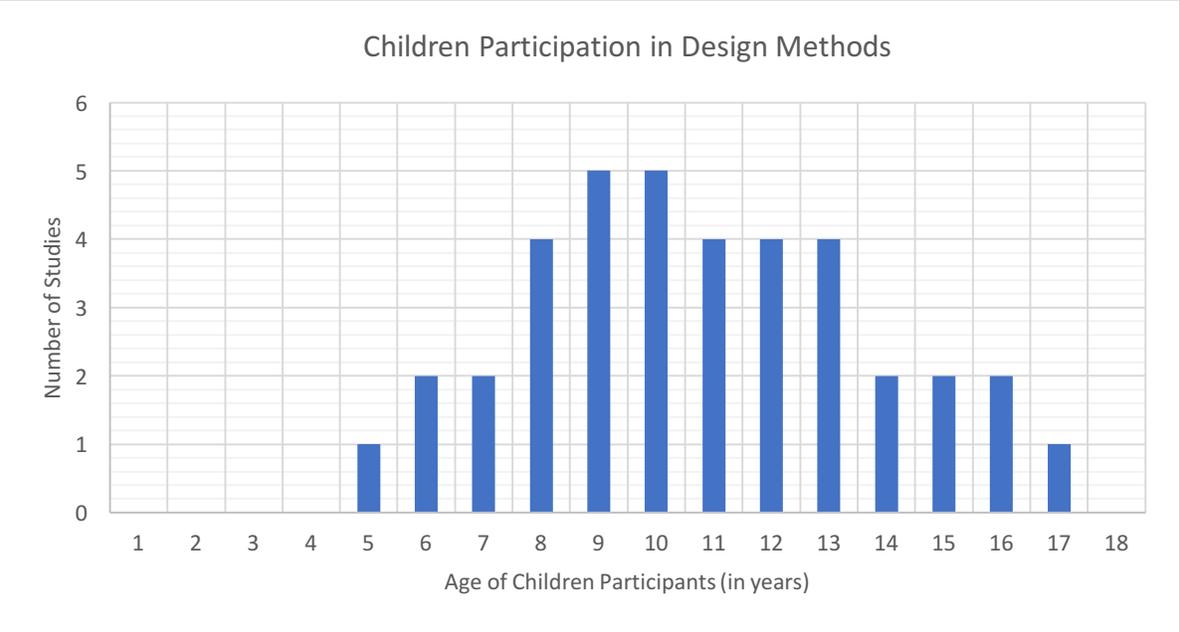

*Figure 4 Number of studies based on the age of children participated in the design methods*

*Table 4 Papers distribution based on themes and target populations*

| No. | Theme | Adult with ADHD | Children with ADHD | Caregiver/ Parents | % (N=27) |
|---|---|---|---|---|---|
| 1 | Attention deficit | Beaton et al., 2014 | Hansen et al., 2017<br>Sonne et al., 2015 | Zuckerman et al., 2016 | 15%(4) |
| 2 | Executive functioning | Huh & Ackerman, 2010 | Eriksson et al., 2017<br>Weisberg et al., 2014<br>Zuckerman et al., 2015 | | 15% (4) |
| 3 | Framework and protocol | Sonne et al., 2016c<br>Zuckerman, 2015 | Sonne et al., 2016c<br>Antle, 2017<br>Asiry et al., 2015<br>Frutos-Pascual et al., 2014<br>McLaren & Antle, 2017<br>Ravichandran and Jacklyn, 2009 | Goldman et al., 2014<br>Sonne et al., 2016c | 30% (8) |
| 4 | Learning deficit | | Kang et al., 2007 | | 4% (1) |
| 5 | Mixed challenges | Dibia, 2016<br>Flobak et al., 2017 | Hashemian & Gotsis, 2013<br>Mandryk et al., 2013 | | 15% (4) |
| 6 | Sensory processing disorder | | Duvall et al., 2016 | | 4% (1) |
| 7 | Sleep disorder | | Sonne et al., 2016a<br>Sonne et al., 2016b | | 8% (2) |
| 8 | Stress release | | Sonne & Jensen, 2016a<br>Sonne & Jensen, 2016b | Pina et al., 2014 | 11% (3) |

3.3.3 Design Methods

After reading each paper, design methods used during requirements, design, and evaluation phases were identified. The results of the design methods used among included literature during requirements phase show that most of the studies relies on findings from previous studies in defining the requirements for their studies (33%), followed by focus group (22%), interviews and questionnaires (17%), then creative sessions and observations (6%). Figure 5 shows the percentages of each design method during requirements phase.

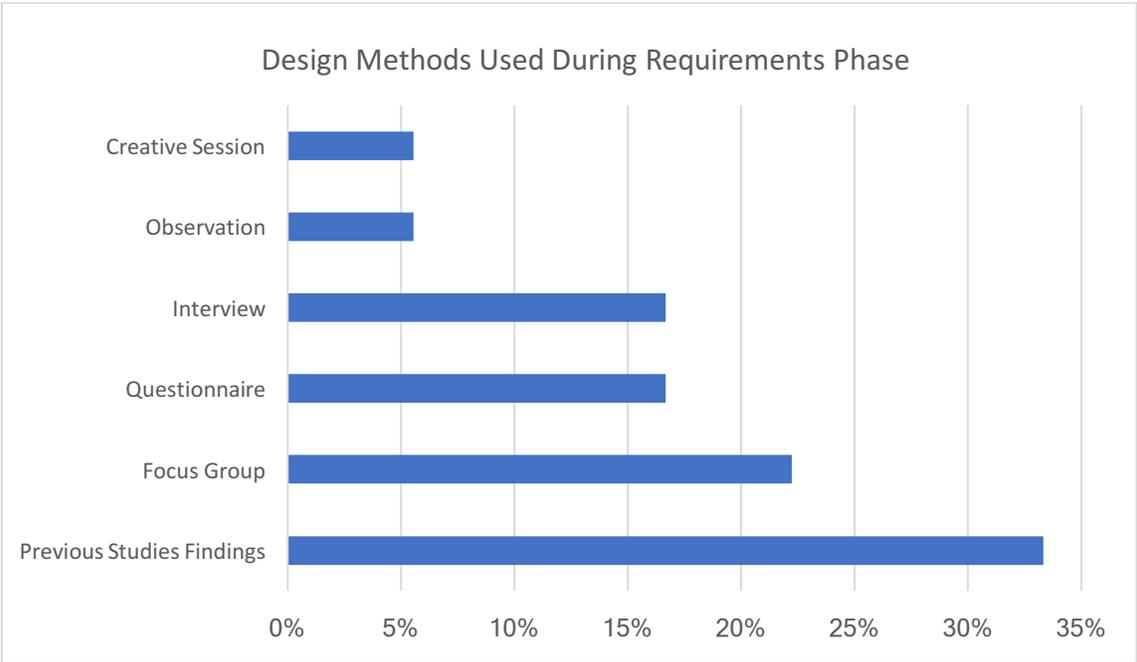

*Figure 5 Percentages of design methods used during requirements phase*

The results of design methods used among included literature during design phase show that most of the researchers used prototyping (e.g., software prototype) in designing their solution (57%), followed by creative sessions (17%), brainstorming and interviews (9%), then design guidelines and field studies (4%). Figure 6 shows the percentages of each design method during design.

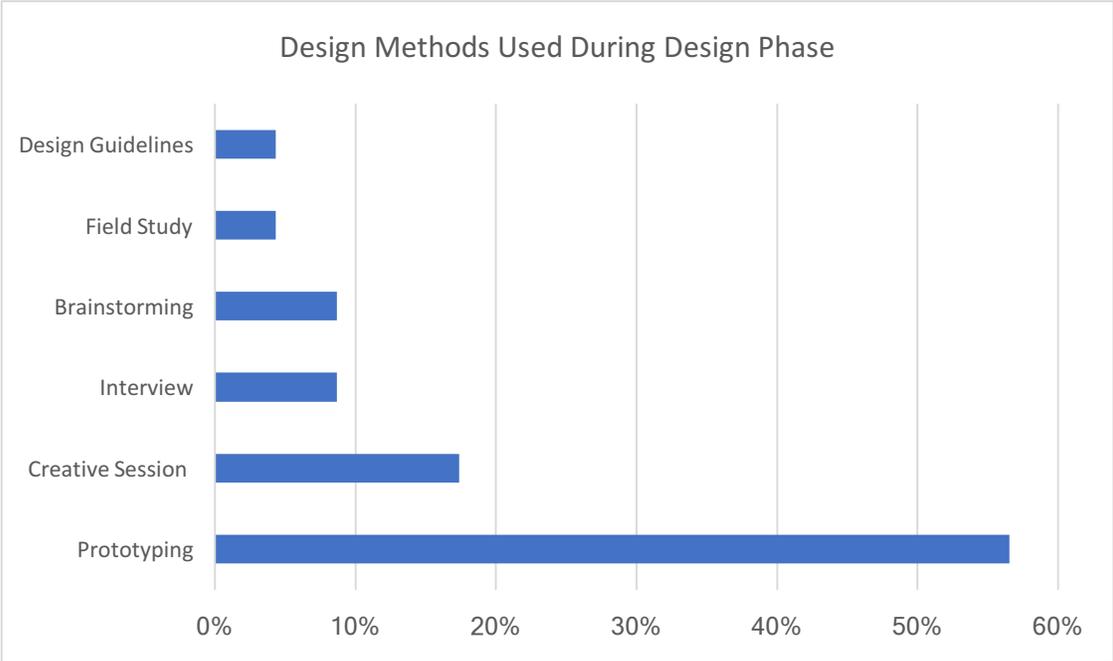

Figure 6 Percentages of design methods used during design

The results of design methods used among included literature during evaluation phase show that most of the researchers used user testing (48%), followed by interview (22%), questionnaire (17%), then cognitive walk through, focus group and observation (4%). Figure 7 shows the percentages of each design method used during evaluation.

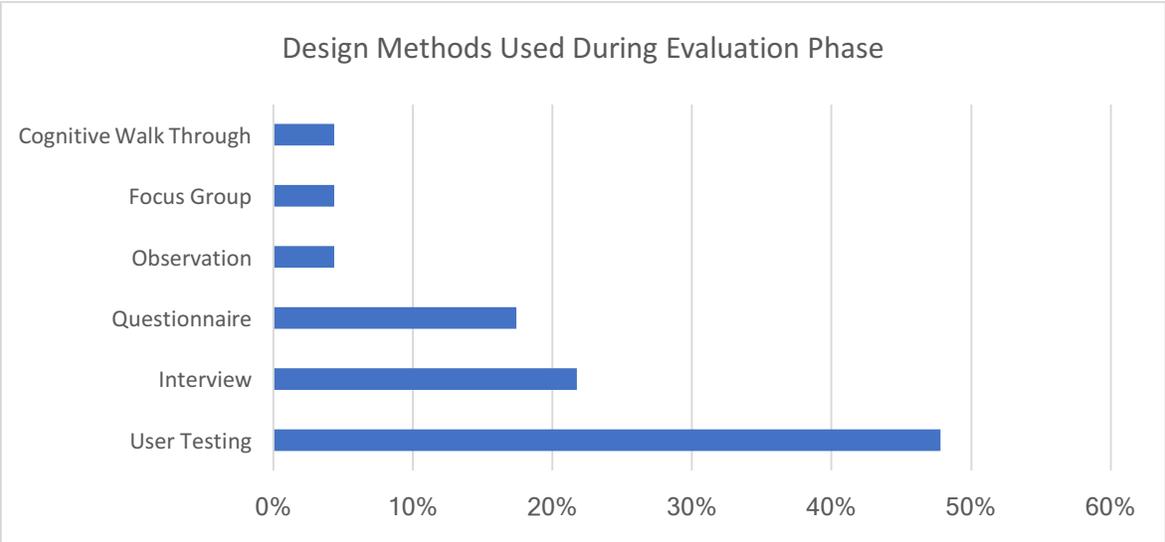

Figure 7 Percentages of design methods used during evaluation

An example of creative sessions found among the methods, one developed by Eriksson et al. (2017), as the authors named it future workshop, has three stages: critique, fantasy, and implementation. Sessions include brainstorming individually and in group, paper prototyping for the solution, and at the end creating an affinity diagram to summarize the findings of the session. Another example of creative session developed by Flobak et al. (2017) to collect requirements from end users in which they establish user panels, including 3 adults with ADHD as repetitive of experts for conditions and needs of individuals with ADHD, each user panel has multiple design workshop to encourage discussion and knowledge sharing among participants to generate innovative design ideas.

Prototyping was the dominant method among the literature for the design phase. Some researchers used more than one prototyping methods either because their system has multiple interface platforms or for testing user preferences. In most cases, the prototype involves an initial design for the system with the main features mainly for concept testing. Among the prototyping methods used during designing phase were software prototype, such as mobile application prototype (Sonne et al., 2016a; Beaton et al., 2014; Pina et al., 2014; Flobak et al., 2017) or tablet application (Pina et al., 2014; Zuckerman et al., 2015; Zuckerman et al., 2016). Other researchers use tangible object prototype (Hansen et al., 2017; Sonne & Jensen, 2016; Weisberg et al., 2014 and Zuckerman et al., 2015; Zuckerman, 2015), wearable prototype (Sonne et al., 2015; Eriksson et al., 2017; Dibia, 2016; Duvall et al., 2016), social robotic (Zuckerman et al., 2016), video game (Sonne & Jensen, 2016; Hashemian & Gotsis, 2013; Goldman et al, 2014) and paper prototype (Eriksson et al., 2017; Weisberg et al., 2014). Web based prototype was not used in any study.

Findings on questionnaires method show that some researchers use domain specific questionnaires as measurements to assess the results of the study. For example, in sleep disorder studies, the Children Sleep Habit Questionnaire (CSHQ) used to be filled by parents of children with ADHD as a pre-test and post-test to evaluate the efficacy of the system (Sonne et al., 2016a).

An example of focus group used within requirements phase for designing personalized fidget for ADHD student involves university researchers, special education teacher, & staff at science museum to ensure accurate identification of requirements (Hansen et al., 2017).

Previously defined design guidelines and protocols were used in some studies. Examples of such design guidelines include Saxon's textbook to generate test materials to measure student's math skills (Kang et al., 2007) and Barkley's principles for designing assistive technology (Zuckerman et al., 2016), which were used during the design phase in both studies.

3.3.4 Contexts of Use

The solutions proposed by the included literature were designed to be used in three main settings: home (63%), school (16%), both home and school (21%). Clinical specific solution does not exist within the included literatures. Figure 8 shows the percentages of the context of use.

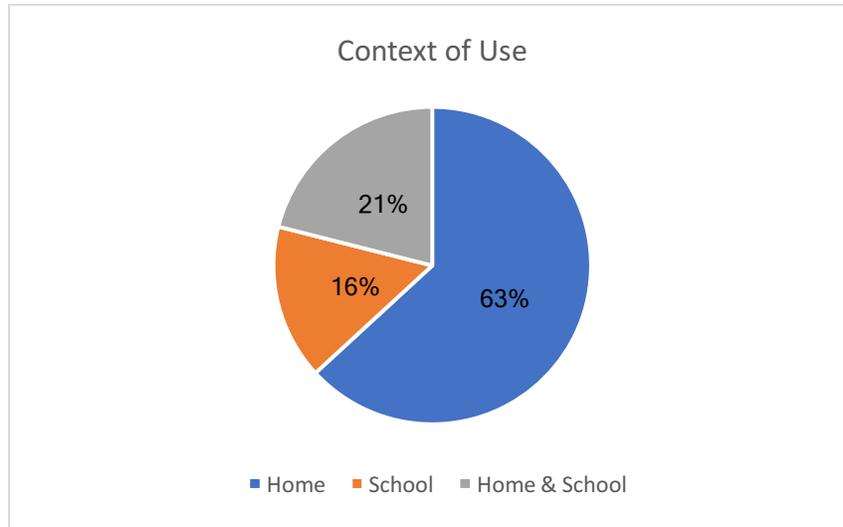

*Figure 8 Percentages of Context of use found from papers*

3.3.5 Sample Participants Size

Evaluating solutions using user-centered design methods might be very costly in term of efforts and time to achieve and analyze. Consequently, it is usual to have small sample size among design methods. Findings on children with ADHD, show most studies involve 12 to 13 participants, 2 participants were the smallest sample size and 16 participants was the largest sample size. Figure 9 shows detail on children sample sizes found among studies.

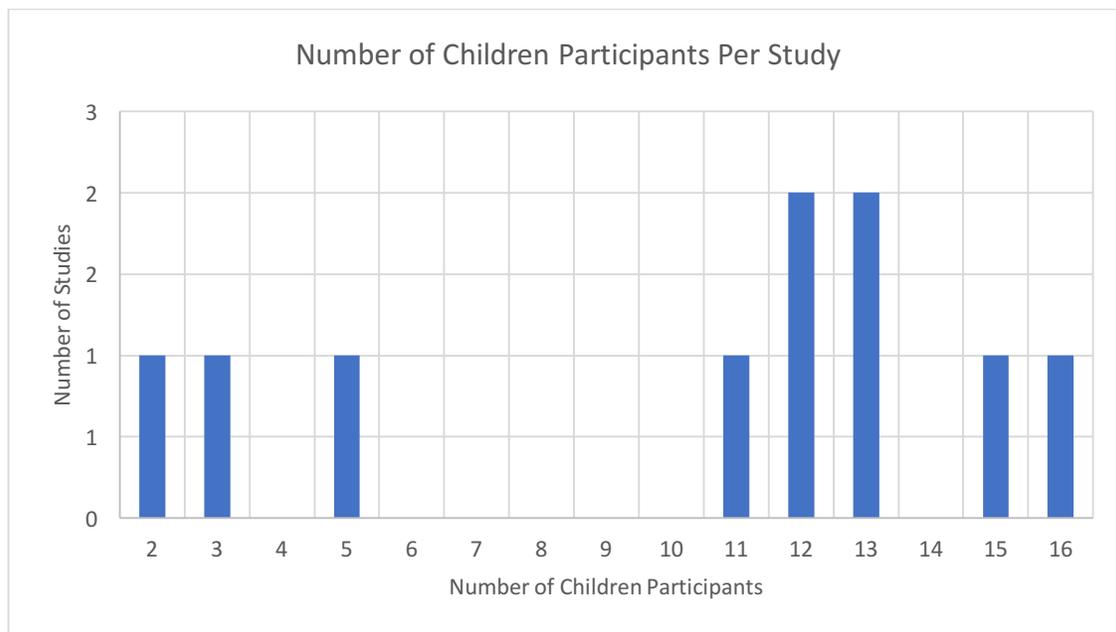

*Figure 9 Number of children with ADHD participants per study*

Findings on adults with ADHD including caregivers and parents of children with ADHD, show most studies involve 10 participants, 6 participants were the smallest sample size and 16 participants was the largest sample size. Figure 10 shows details on adult's sample sizes found from studies.

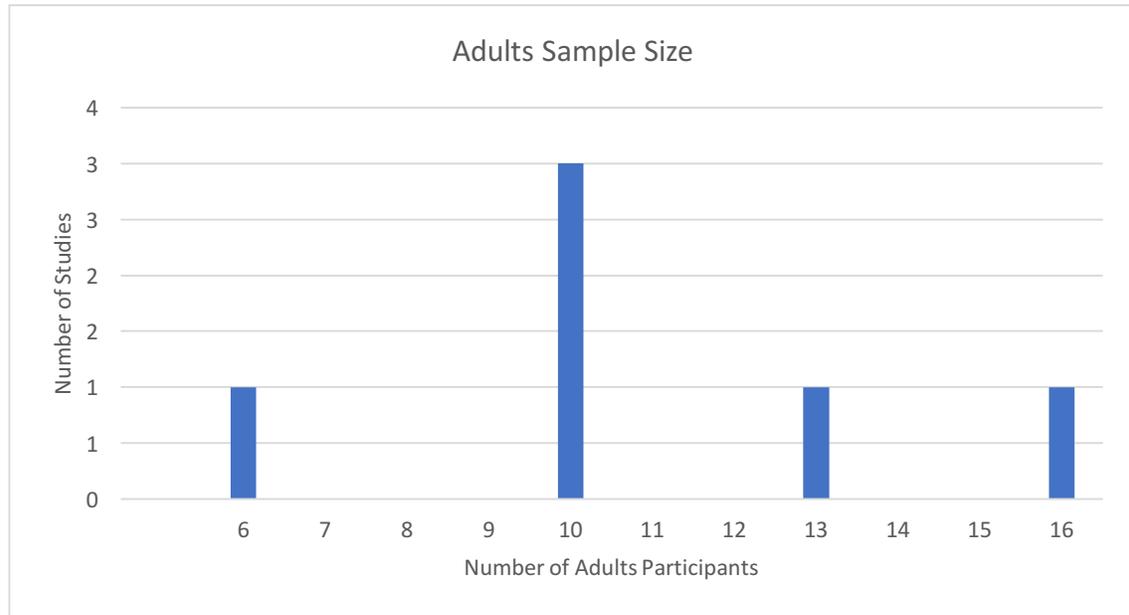

*Figure 10 Number of adults with ADHD participants per study*

3.3.6 Overview of Literature Involving Design Methods Based on Challenge

Sleep Disorder:

Among the papers, one sleep assistive tool was found, and two studies evaluated its impacts. MOBERO is a mobile behavior intervention system aimed to establish a morning and bedtime routines for children with ADHD to reduce symptoms related to sleep disorder. The application was evaluated during baseline, intervention, and follow-up periods, and the results of all evaluations show positive changes in children sleeping behaviors (Sonne et al., 2016a; Sonne et al., 2016b). Table 5 shows details on sleep disorder papers.

*Table 5 Sleep disorder papers*

| Study | Technology/Study Description | Evaluation Participants | Design Method | Targeted Population | Result | Limitations |
|---|---|---|---|---|---|---|
| Sonne et al., 2016a | Designing and evaluating a mobile system to assist families of children with ADHD in establishing healthy morning and bedtime routines: MOBERO | 11 families and 13 children with ADHD (4 female, 9 male) ages 6-12 (average 9.3) | **Requirements:** questionnaires **Design:** prototyping (concept testing) **Evaluation:** semi-structured interview and questionnaire | Children with ADHD – Home | Reduction in parent frustration level and improvement in parent rated child independence level and the child's sleep habits | low number of participants and absence of control condition |
| Sonne et al., 2016b | A follow-up study to assess the long-term effects of MOBERO | 13 children with ADHD and their families | **Evaluation:** interview and questionnaire | Children with ADHD– Home | Fewer conflicts and lower frustration levels around morning and bedtime routines | Conducting the experiment during school summer holiday caused some problems in recruiting participants and effecting family sleep practices |

Attention Deficit:

Four studies found among the papers focus on attention deficit. Their results show that maintaining attention through self-monitoring engagement level in activities along with contextual data on temporal and geographic data can support older individuals with ADHD to better understand their personal behavior (Beaton et al., 2014). Self-monitoring is considered one of the most effective behavior change technique because it increases personal accountability and self-awareness (Zuckerman, 2015).

Three different types of solution were proposed: wearable system, tangible object, and social robotic. CASTT is a wearable real-time assistive system aims at school context use and depends on multiple sensors to detect moment of distraction, and prompt the student using a mobile application to regain attention (Sonne et al., 2015). Fidget is a small hand-held tool that is widely used by students to assist them to maintain focus while in the classroom (Hansen et al., 2017). Hansen et al. (2017) incorporate the benefits of using the fidget with the concept of the Maker Movements (Do-It-Yourself culture) to better suit the needs of students with ADHD to keep focus while at school. KIP3 is a social robotic companion designed to provide real-time cues as a feedback to help adults with ADHD to regain focus for inattention and impulsivity events based on two sets of design guidelines: Barkley's principles; including the use of external information and cues around the person, keep the information within users' sensory fields, and keep cues in the natural environment; and Empathy Objects guidelines; including tangible representation of digital information to supplement human-human interaction. The robot was evaluated with undergraduate students at a university lab using a tablet based Continuous Performance Test (CPT), and the results show its positive effects on student's attention level (Zuckerman et al., 2016). Table 6 shows detailed information of papers on attention deficit.

*Table 6 Attention deficit papers*

| Study | Technology/Study Description | Evaluation Participants | Design Method | Targeted Population | Result | Limitations |
|---|---|---|---|---|---|---|
| Beaton et al., 2014 | Designing a mobile application engagement monitoring tool using an engagement index that uses unobtrusive objective (EEG) and subjective information (e.g., calendar data, mobile GPS). | Not Evaluated | **Requirements:** previous study findings **Design:** creative session and prototyping (software prototype) | College age students with ADHD and ADD – School and Home | Youth with ADHD/ADD maintain their attention by monitoring their engagement level on activities based on temporal and geographic data. | The tool is still in the planning phase, thus needs further investigation to better understand its validity |
| Hansen et al., 2017 | Designing personalized fidget by student with ADHD to better suit their needs to stay focus at school | 5 children with ADHD (3 males, 2 females) $6^{th}$ and $7^{th}$ grades ages 12 (3 students) and 13 (2 students) | **Requirements:** focus group (university researchers, special education teacher, & staff at science museum). **Design:** creative session, interviews (student & teacher) and prototyping (wizard-of-Oz) **Evaluation:** user testing | Children with ADHD- School | Student with ADHD have unique problems, letting them to design the solution produce the optimal solution | Doing the study at the end of the academic year made it hard to collect data on student attention in class while using their fidgets to determine its effectiveness |

| Study | Technology/Study Description | Evaluation Participants | Design Method | Targeted Population | Result | Limitations |
|---|---|---|---|---|---|---|
| Sonne et al., 2015 | CASTT: a wearable real-time assistive system | 11 children with ADHD and 9 children without ADHD ($3^{rd}$, $4^{th}$ or $5^{th}$ grade) | **Requirements:** observation (school) and interviews (professionals) **Design:** prototyping (concept testing) **Evaluation:** observation (school) and biofeedback (EEG and HR) | Children with ADHD- School | Comfortable to wear. A smartphone-based intervention assist to remain focus | EEG headset has some wearing problems to children (size fitting) |
| Zuckerman et al., 2016 | Designing KIP3, a social robotic companion to provide real-time feedback for inattention | 10 undergraduate students (4 males, 6 females) ages 20-35 (average 26.3) | **Requirements:** previous study findings **Design:** design guidelines **Evaluation:** user testing and interview (semi-structured) | Adults with ADHD – School and Home | Providing real-time cues by a social robotic helps adults with ADHD to regain focus. | Concerns on long-term effects outside the lab. Reaction of others around the users to the robot |

Executive Functioning:

Executive Functioning (EF) skills cover multiple activities that involve self-regulation such as time management and goal achievement. Lacking EF skills creates more severe problems for adults than children with ADHD because they have more complicated issues to deal with in their everyday life such as higher education, career duties, and families. Huh & Ackerman (2010) investigate in depth the needs and requirements of adults with ADHD in learning Personal Information Management (PIM) strategies from their social environment as a method to support EF skills.

Among the studies, two tangible systems aim to support children with ADHD in EF skills, TangiPlan and Takt. TangiPlan is a tangible interface system connecting objects that represent morning activities to help middle school children with ADHD to perform EF tasks, and it is connected with a web-based interface and analytical tool to support real-time monitoring (Weisberg et al., 2014). Evaluation of TangiPlan shows its positive feedback on daily morning activities, but the authors highlighted some technical problems with using battery-based devices (Zuckerman et al., 2015). Takt assists school aged children with ADHD in managing their time through a haptic feedback (continuous vibration) and visualizing the activity as a time duration rather than a point in time using a tangible system (Eriksson et al., 2017). Table 7 contains detailed information on papers involving EF skills.

*Table 7 Papers on executive functioning*

| Study | Technology/study Description | Evaluation Participants | Design Method | Targeted Population | Result | Limitations |
|---|---|---|---|---|---|---|
| Eriksson et al., 2017 | Designing Takt a wristband time management haptic feedback system that focus on activity-based visual representation rather than point in time | 15 children with ADHD ages 13-16 | **Requirements:** previous study findings and creative sessions. **Design:** brainstorming, prototyping (paper prototype, concept testing, & 3D printed prototype), and creative session. **Evaluation:** focus group (students & researchers) | School age children with ADHD – Home and school | Students want different vibration intervals and colors for different events, and they want to control the system through direct touch rather than an external app | Not Mentioned |

| Study | Technology/study Description | Evaluation Participants | Design Method | Targeted Population | Result | Limitations |
|---|---|---|---|---|---|---|
| Huh & Ackerman, 2010 | Understanding the social dimension of Personal Information Management (PIM) behavior of adults with ADHD for better design implications | Requirements Participants: 13 adults with ADHD and 3 coaches | **Requirements:** interview (semi-structured) | Adults with ADHD – Home | Adults with ADHD seek help to adapt with PIM strategies from multiple sources (e.g., experts, parents, online forums) and by using multiple iterative practices learned from their social environment | Lacking a specialized social sharing tool for PIM strategies |
| Weisberg et al., 2014 | TangiPlan: a system of tangible connected objects representing morning time tasks connected with web-based interface and analytical tool | Design Participants: 3 children with ADHD (all males) and their parents (all females) | **Requirements:** interviews (professionals and users). **Design:** prototyping (paper prototyping and 3D tangible concept testing prototype) and interviews (children and parents) | Middle school children with ADHD- Home | Tangible system is less distracted than mobile system and can be associated with location specific functions | Not Mentioned |

| Study | Technology/study Description | Evaluation Participants | Design Method | Targeted Population | Result | Limitations |
|---|---|---|---|---|---|---|
| Zuckerman et al., 2015 | Validation of TangiPlan | 2 children with ADHD (1 male, 1 female) ages 13 and 13.5 | **Evaluation:** questionnaire (children and parents) and semi-structured interviews (children and parents) | Middle school children with ADHD- Home | Increase in satisfaction with morning routine and reduction in parental involvement | Technical problems related to battery depletion and WiFi connection |

Learning Deficit:

Kang et al. (2007) emphasize on the great opportunities of integrating technology into the instructional curriculum at school, specifically for subjects that need attention such as geometry for children with ADHD. Table 8 shows detailed on this study.

*Table 8 Papers on learning deficit*

| Study | Technology/Study Description | Evaluation Participants | Design Method | Targeted Population | Result | Limitations |
|---|---|---|---|---|---|---|
| Kang et al., 2007 | Utilizing images to provide stimulation for students with attentional problems and direct their attentional function in learning geometry | 2nd to 4th grade students with and without characteristics of ADHD | **Design:** field study (math teachers and Saxon's textbook) **Evaluation:** prototyping (software) and pre-test and post-test measurements | Children with ADHD – School | The use of Additional Visual Cues (AVC) images were more helpful than Low-level Visual Cues (LVC) images in learning difficult geometry terminology and calculations of perimeters | Not Mentioned |

Sensory Processing Disorder (SPD):

Individuals with ADHD are more likely to experience anxiety and depression as a result of their Sensory Processing Disorder (SPD), especially for children, thus finding some ways to provide Deep Touch Pressure (DTP) support them as a tool for swaddle therapy which is one of the treatments of SPD. Duvall et al. (2016) designed a smart vest with three main components: electronics system (to receive signals from smartphone and activate SMA network), Shape Memory Alloy (SMA) actuators (have multiple states to support changing from one to another), and physical prototype (3 layers garment: inner layer, muscle layer, and outer layer). Table 9 shows detailed on this study.

*Table 9 Papers on sensory processing disorder*

| Study | Technology/Study Description | Evaluation Participants | Design Method | Targeted Population | Result | Limitations |
|---|---|---|---|---|---|---|
| Duvall et al., 2016 | Designing a smart vest with Shape Memory Alloy (SMA) to provide swaddle therapy through Deep Touch Pressure (DTP) using a mobile application to support adjustability and controllability by the user or others | Not Evaluated | **Requirements:** previous study findings | Children with ADHD – School and Home | Design consideration to providing a vest with DTP for swaddle therapy, should have three main components: 1) electronics system (to receive signals from smartphone and activate SMA network), 2) SMA actuators (have multiple states), and 3) physical prototype (3 layers garment) | Needs further investigation with experts and occupational therapists |

Mixed Challenges:

Among the papers, two studies focused on children with ADHD to support them on more than one challenge. Hashemian & Gotsis (2013) developed the Adventurous Dreaming Highflying Dragon prototype for a full-body driven game for children with ADHD between the ages of 5 and 8, incorporating physical exercises and play-oriented interventions to improve attention, motor skills, and EF, using audio feedback rewarding system. Mandryk et al. (2013) designed a biofeedback training system to support children with Fetal Alcohol Spectrum Disorder (FASD) and ADHD to self-regulate their brain function, the main idea of their prototype is to have a multiple textural-based graphical overlays that offer flexibilities to the user through customization of the overlay design, sensor types and skills to be trained, and being an off-the shelf system that can be used on top of any game, web browser, or software.

Another two studies focused on adults with ADHD supporting them by adapting with multiple challenges. Flobak et al. (2017) design a customized mobile application that depends on Patient-Generated Data (PGD) as an evidence-based approach through a wristband device to sense Electro Dermal Activity (EDA) and other psychophysiological data, and to provide in-situ intervention. The second study proposes FOQUS a smartwatch application for adults that support maintaining their attention and reducing stress. FOQUS receives user EEG and heart rate, and based on both sensing data provides messages (positive messages and health tips), and delivers Promodoro time management technique to support attention and breathing meditation to support stress release (Dibia, 2016). Table 10 shows details on literatures involving more than one challenge.

*Table 10 Papers involving more than one challenges*

| Study | Technology/Study Description | Evaluation Participants | Design Method | Targeted Population | Result | Limitations |
|---|---|---|---|---|---|---|
| Dibia, 2016 | Designing FOQUS a smartwatch application to support extending attention through Promodoro time management technique, and reduce stress through visual and haptic cues to regulate breathing pattern, both implemented with either positive messages or health tips | 10 participants ages 21-30 | **Requirements:** survey (user) **Design:** prototyping (software) **Evaluation:** cognitive walkthrough and usability tests | Adults with ADHD | Smartwatch-based intervention has positive effects on extending attention when functions were grouped into multiple screens, and the breathing exercise has positive effects on reducing stress. Users have some concerns regarding the accuracy of the sensing data and the context of use | Small number of participants and not including measurements for the efficacy of the system |

| Study | Technology/Study Description | Evaluation Participants | Design Method | Targeted Population | Result | Limitations |
|---|---|---|---|---|---|---|
| Flobak et al., 2017 | Designing an assistive technology to train cognitive and emotional control skills through Patient-generated data (PGD) as a pervasive affective sensing to support affective interactions in a well-documented treatments approach | 6 participants | **Requirements:** previous study findings **Design:** creative session (users) and prototyping (mobile application and wristband porotype) **Evaluation:** user testing | Adults with ADHD | Not Mentioned | Not Mentioned |
| Hashemian & Gotsis, 2013 | Demonstration of the Adventurous Dreaming Highflying Dragon game which incorporate physical activities to improve ADHD symptoms related to attention, executive functioning, and motor skills | 12 children with or without ADHD ages 5-8 | **Requirements:** focus group (faculty researchers and clinicians) **Design:** prototyping (software prototype and concept testing) **Evaluation:** informal user testing | Children ages 6-8 with ADHD – Home | Not Mentioned | Not Mentioned |

| Study | Technology/Study Description | Evaluation Participants | Design Method | Targeted Population | Result | Limitations |
|---|---|---|---|---|---|---|
| Mandryk et al., 2013 | Off-the-shelf customized texture-based biofeedback graphical overlays | 16 children with FASD (9 male, 7 female) ages 8-17 (median=11) | **Requirements:** previous study findings **Design:** prototype (software prototype and concept testing) **Evaluation:** user testing (end users), biofeedback (EEG log files), and survey | Children with ADHD and FASD-Home | The texture-based biofeedback overlay did not ruin the fun of games for children indicating the potential of supporting users to self-regulate their brain function | Not Mentioned |

Stress Release:

Among the included papers, two studies with stress relief purpose were found, one for children with ADHD and another for parents of children with ADHD. ChillFish is a calming system that helps children with ADHD to control their stress levels through a biofeedback breathing exercise game (Sonne & Jensen, 2016a; 2016b). The child uses a LEGO fish as a breath-based respiration controller for ChillFish game, a sensor detects the child's HRV values as a quantitative stress indicator, the location of the fish in the video game is based on HRV values, and the player catches starfishes which are arranged in a way to support healthy breathing exercise (Sonne & Jensen, 2016a; 2016b). Results of evaluating ChillFish show some technical problems with using a physical sensor-based device with children, that did not occur when the system was evaluated by adults, such as children exhale an extended amount of saliva into the LEGO fish causing the electronics to stop working and children found it hard not to fiddle with the EDA electrodes, which makes the sensor values changed, making the EDA data invalid, thus HRV value was used as a stress indicator (Sonne & Jensen, 2016b). ParentGuardian is a mobile and peripheral system to be used for parents of children with ADHD, developed based on a wearable EDA sensor to detect the stress level on a real-time and provide in situ strategies for Parental Behavioral Therapy (PBT) on mobile screen (textual PBT and visual PBT) and peripheral screen (visual PBT) to help parents in stress relief (Pina et al., 2014). The results of evaluating ParentGuardian highlight the effects of having reliable real-time cues, and timing and duration of those cues on positive PBT. Table 11 shows details of these two studies

*Table 11 Papers on stress release*

| Study | Technology/Study Description | Participants | Design Method | Targeted Population | Result | Limitations |
|---|---|---|---|---|---|---|
| Pina et al., 2014 | ParentGuardian: a mobile/peripheral system provides in situ cues for ADHD Parental Behavioral Therapy (PBT) in moments of stress | 10 parents of children with ADHD with average age of 38.4 (8 females, 2 males) | **Requirements:** questionnaire and focus group **Design:** brainstorming and prototyping (concept testing) **Evaluation:** user testing (biofeedback- EDA) and post experience interviews | Parents of children with ADHD – Home | In situ cues have positive effects on supporting PBT during stress times, but it is very important to have reliable cues and within reasonable timings considering the escalation of stressful situation | The wearable EDA sensor is limited to Bluetooth connection and does not support user interaction through screen. Machine learning algorithm was not trained with actual users and thus results were not matched |
| Sonne & Jensen, 2016a | ChillFish: a calming biofeedback game designed for children with ADHD by combining a breathing exercise with a video game. | 16 adults aged 25-41 (14 males, 2 females) | **Requirements:** focus group **Design:** prototyping (concept testing) **Evaluation:** user testing (causal conversation and biofeedback- HRV) | Children with ADHD - Home | Combining breathing exercise with game design reduce stress level. Game elements and theme impact the calming experience. Game dynamics facilitated reflective experiences | Low number of participants. The use of HRV as a measure for stress which are affected by the respiration of the subject, and hence breathing exercises might have an effect |

| Study | Technology/Study Description | Participants | Design Method | Targeted Population | Result | Limitations |
|---|---|---|---|---|---|---|
| Sonne & Jensen, 2016b | Evaluating the ChillFish biofeedback game with children with ADHD | 12 children with ADHD aged 8-13 | **Evaluation:** user testing (interview and biofeedback-HRV) | Children with ADHD – Home | HRV result indicates a positive tendency that playing ChillFish and performing a traditional breathing exercise provides a similar calming experience for children | Children exhale an extended amount of saliva into the LEGO fish causing the electronics to short. Children found it hard not to fiddle with the EDA electrodes, thus sensor values changed, making the EDA data invalid |

### 3.3.7 Overview of Design Framework and Protocol

Many of the studies included (30%) focused on design guidelines and protocols for systems supporting individuals with ADHD providing frameworks for developers and future researchers. Sonne et al. (2016c) provide guidelines for designing strategies for HCI research within the ADHD research based on investigating the existing empirical studies, ADHD research, and related work on assistive technologies. Researchers emphasized on the importance of four strategies: providing structure to facilitate activities as people with ADHD are more likely to complete tasks when they are presented in a predictable pattern (e.g., chart, checklist), minimize distractions to help in extending their attention (e.g., seating the child in the front of the class), encourage praise and rewards especially for children and teenagers, and integrate and report standardized ADHD measures to maximize the impact of the study (measures should include severity of ADHD symptoms, medical treatments during the evaluation periods, and qualitative findings such as unexpected outcome and unique challenges).

Ravichandran and Jacklyn (2009) focus on children below 12 years old with attention deficit for educational purposes. They propose a strategy to support problem solving skills through an audio-visual animation rewarding system. Frutos-Pascual et al. (2014) narrow down the protocol to be specific for cognitive rehabilitation interventions using serious games and biofeedback techniques for children with ADHD. They provide detailed descriptions on the materials and methods that should be used in validating the evaluation of such techniques. Moreover, their protocol includes six minimum criteria: defining standards for performing the experiment, defining instruments and validation procedures, having standard intervention environment based on specialized children therapists recommendations, pre-test and post-test validation measurements (e.g., questionnaires, scales), randomly divide participants into two groups: standard intervention participants and extended intervention participants, and finally statistically quantify changes between the results of both groups.

Framework including design principles that utilize the procedural rhetoric strategies in educating people about ADHD show positive effects in increasing ADHD awareness among non-ADHD adults (Goldman et al., 2014). Drawn to Distraction is a persuasive video game aims to facilitate understanding of ADHD, the results of its evaluation show positive effects of using the procedural rhetoric strategies to promote understanding of the symptoms and the difficulties that are facing individuals with ADHD by adults without ADHD (Goldman et al., 2014).

A Theoretical framework proposed by McLaren and Antle (2017) emphasizes on the importance of using sound treatments (e.g., binaural beats, white noise) as a background sound for the lessons to improve the attention of children with ADHD at school.

Asiry et al. (2015) propose a framework for a computer software system aimed to extend the attention span for school age children with ADHD through eye tracking. The design framework is based on detecting the attention state of the child through the eye tracker, and adjust the system interface to keep the child focused on the tasks (solving mathematic problem) for longer time through changing the colors (e.g., use warm colors for on-task information), highlighting on-task information (e.g., blurring the off-task information), and converting text to speech to bring the child back to the task whenever the child get distracted (Asiry et al., 2015).

Antle (2017) argues on the difficulties of doing research with children particularly on ethical challenges. She opens discussion on a set of issues that should be considered within the design framework, such as whether the children are benefiting from the research, reporting the results carefully regarding the involved children, considering that generalization not always right, and considering the differences in cultures. These ethical considerations create positive effects on the lives of children who are the targeted users of the systems.

Developing persuasive technology with Tangible User Interface (TUI) has unique design prosperities differ from other non-tangible interface technologies. Zuckerman (2015) proposed four main properties for designing behavioral change TUI for youth with ADHD: (1) visibility and persistency (part of daily ritual and single purpose), (2) locality (location-dependent properties), (3) tangible representation (e.g., physical movement, sound), and (4) affordance (change users experience from perception to reflection). Table 12 shows detailed information on papers involving design framework and protocol.

*Table 12 Papers involve design framework and protocol*

| Study | Targeted Population | Research Method | Result | Limitations |
|---|---|---|---|---|
| Antle, 2017 | All children | Ethical and cultural challenges of doing research with children | Children involved in the research should understand the research, ensure that the children are benefiting from the research, reporting the results of children involved research should be considered carefully, generalization not always right, and consider the differences in cultures | Not Mentioned |
| Asiry et al., 2015 | School age children with ADHD | Investigating existing research on improving attention for children with ADHD to design a software system that can adapt its interface to extend the attention span based on multiple stimuli for children with ADHD at school | Using the eye tracker to detect child attention state and adapt the users interface adjusting the colors, highlighting tasks, and converting text to speech can extend the attention span measured by the time spent on task and the accuracy of the results | Not Mentioned |
| Frutos-Pascual et al., 2014 | Children with ADHD | Inclusion of biofeedback and serious game techniques within intervention program | The minimum criteria for validating experiment on rehabilitation intervention using biofeedback and serious games are defining standards, defining instruments and validation procedures, having standard intervention environment, having pre-test and post-test validation measurements, having standard intervention participants and extended intervention participants, and statistically quantify changes | Professionals must be trained on the intervention patterns because the protocol aimed at children |

| Study | Targeted Population | Research Method | Result | Limitations |
|---|---|---|---|---|
| Goldman et al., 2014 | Adults without ADHD | Feasibility study of a prototype for persuasive video game "Drawn to Distraction" aims to facilitate understanding of ADHD through procedural rhetoric framework | The use of serious persuasive games through procedural rhetoric shows positive effects on understanding of ADHD | Results are not conclusive because of the small sample size, and the questionnaire used in the measurements focused on factual-knowledge and missing empathy-based measurements |
| McLaren & Antle, 2017 | Children with ADHD | Evaluating the effect of auditory treatments on school (binaural beats and white noise) on improving children's attention and self-regulate through a neurofeedback system "MindFull" | A clear systematic methodology is needed to better understand how sounds can influence neurofeedback training, having both objective and subjective measures of attention is very important | No follow-up study to validate the reliability of the framework, and sampling issues such as small sample size and lacking randomization. |
| Ravichandran and Jacklyn, 2009 | Children with ADHD | Investigating the existing behavior modification strategies for problem solving skills | The use of audio-visual animation as a rewarding system for children increase their attention | The success of the strategy depends on the level of motivation in learning and reward expectation of subjects |

| Study | Targeted Population | Research Method | Result | Limitations |
|---|---|---|---|---|
| Sonne et al., 2016c | Parents, adults and children with ADHD | Investigating empirical studies, ADHD research, and related work on assistive technology to provide HCI researchers with design strategies | There are no existing assistive technologies that capture contextual data for later retrieval or that focus on risky behaviors (e.g, accidents). Four strategies for HCI research: provide clear structure, minimize distractions, use praise and rewards, and integrate standardized ADHD measurements | Not Mentioned |
| Zuckerman, 2015 | Youth with ADHD | Providing design guidelines for Tangible User Interface (TUI) of persuasive behavior change technologies based on theoretical framework, interviewing experts, caregivers, and individuals with ADHD, and prototyping | Designing TUI to support behavior change should have four properties: visibility and persistency (part of daily ritual and single purpose), locality (location-dependent properties), tangible representation (e.g., physical movement, sound), and affordance (change user's experience from perception to reflection) | Prototype is not evaluated to validate the proposed design guidelines |

## 4. DISCUSSION

The aim of this systematic literature review is to investigate the current state of assistive technology literature covering Attention Deficit Hyperactivity Disorder (ADHD). Researchers from Human-Computer Interaction (HCI) usually target children with ADHD, few researchers study caregivers and parents of individuals with ADHD. This domain of research is within its establishment phase, researchers more actively focused on assistive technologies for individuals with ADHD from 2013. The results of the review show that most studies focus on design frameworks and protocols to provide design guidance and procedures for researchers and developers.

The symptoms of ADHD usually show-up on early childhood. Most children get diagnosed by the age of seven. Early support during childhood prevents the manifestation of its symptoms before entering adulthood. Even though most researchers focus on children, there are still valuable opportunities for further research on children with ADHD (Sonne et al., 2016c). Mapping between the current state of the art for research of assistive technology for children with ADHD and the challenges defined by the Children and Adults with Attention-Deficit/ Hyperactivity Disorder (CHADD) organization shows existing gaps in research (see Table 1). Behavioral disorder including conduct disorder and oppositional defiant disorder have not been explored yet, despite its high prevalence (1 out of 4 for conduct disorder and 1 out of 2 for oppositional defiant disorder). Only one researcher focus on learning disorder even though 1 out of 2 children with ADHD are facing learning difficulties which might affect their academic life.

Designing new solutions should utilize the positive findings from previous studies regarding system interface and interact with individuals with ADHD including Tangible User Interface (TUI), audio rewarding system, physical activities involvement, the design of the environment, empathy and emotional interaction. Multiple research highlighted the importance of TUI in changing behaviors (Zuckerman, 2015; Sonne & Jensen, 2016a; Sonne & Jensen, 2016b; Weisberg et al., 2014; Zuckerman et al., 2015). Furthermore, Zuckerman (2015) extends it further and defined four TUI principles: 1) visibility and persistency, 2) locality, 3) tangible representation, and 4) affordance. The use of audio reward system helps keeping the children involved and engaged with the system (Ravichandran and Jacklyn, 2009; Hashemian & Gotsis, 2013). Developing a solution that involves physical activities can improve the ADHD symptoms (Sonne & Jensen, 2016a; Sonne & Jensen, 2016b; Hashemian & Gotsis, 2013). It is critical to consider the main purpose of the system when designing the environment of the system to support the children to stay focused on the main purpose of the system and to ensure consistency among the system components (Sonne & Jensen, 2016a; Sonne & Jensen, 2016b; Hashemian & Gotsis, 2013). The design that provides real-time empathetically and emotional cues to users can keep them stayed focused with the system (Zuckerman et al., 2016).

There are some significant findings that are not directly related to the user interface of the system but are very important to consider, such as having a theoretical basis, and ethical concerns. Coming up with an idea of the solution based on existing social theories gives more strengths and increase its efficacy. Hansen et al. (2017) emphasize on the importance of utilizing the Maker Movement in optimizing the proposed solution by engaging the users (e.g., children with ADHD) in developing a personalized solution or tool to ensure satisfaction. Alissa Antle (2017) raises some ethical concerns regarding the inclusion of children in research such as the end benefits that kids are getting from the study and the integrity of researchers' assumptions on children developments and life. She also argued on the importance of culture specific design and localization of solutions.

Considering the limitations of the current research is very important to avoid having the same mistakes. The use of mobile application can increase distraction (Sonne & Jensen, 2016a; Weisberg et al., 2014), because the user might get exposed to prompts from other mobile applications. Problems in fitting the wearable devices and size issues with children (Sonne et al., 2015; Duval et al., 2016). Children with ADHD are vulnerable users, testing the system with adults before do so with children with ADHD is better to ensure avoiding any control or ethical issues (Sonne et al., 2016a). Avoid recruiting children during the summer holiday or at the end of the academic year because families usually change their daily practice which might affect the accuracy of the results (Sonne et al., 2016b; Hansen et al., 2017). Involving experts and therapists in the study can help in defining better requirements (Duval et al., 2016). Avoid developing a system that requires battery and Internet connection as battery might get depletion during the experiment and Internet might be disconnected (Zuckerman et al., 2015). Avoid having Electro-Dermal Activity (EDA) sensors with children as it is very sensitive to movements and it is hard to control the movements of children (Sonne & Jensen, 2016b; Pina et al., 2014). Use an empathy based measurements to reveal more behavioral information which might be hard to get from questionnaires and to assess users experience with the system more accurately (Goldman et al., 2014). Follow-up study can reveal more information about the long-term effects of the use of the system on participants which is very important to ensure the long-term success of the solution (McLaren & Antle, 2017; Zuckerman et al., 2016).

## 5. LIMITATIONS

The findings in this literature review are limited to ACM DL, and thus they might be biased toward publications in ACM journals and conferences. The analysis has some limitations on demographic background information of participants, most studies include participants' information regarding their age and gender, but none includes race information (e.g., Hispanic, white), or socio-economic level, which have crucial impact on the diagnosis and treatments related to ADHD (Hinshaw et al., 2011). Lacking control group in the studies, among included papers, only three studies have control groups (Hashemian & Gotsis, 2013; Sonne et al., 2015; Kang et al., 2007). Having a control group in the evaluation of the system assesses the accuracy and quality of the results.

## 6. CONCLUSIONS

This literature review highlighted the importance of exploring the research opportunities on assistive technologies for individuals with ADHD. Researchers in Human-Computer Interaction (HCI) should consider the limited studies on ADHD to cover the current gaps in research, and to provide greater support for individuals affected by ADHD.

The contribution of this review can help HCI researchers to identify the procedures and research methods used throughout requirements, design, and evaluation phases in developing assistive technology for individuals with ADHD. Moreover, it provides researchers with information regarding frameworks and protocols of conducting studies on ADHD, current available solutions, and their limitations.

Retrieved on March 20, 2018 from
http://www.chadd.org/Portals/0/NRC/AboutADHD/PDF/Infographic3.pdf .

National survey of children's health (2017, September 6). Centers for disease control and prevention. Retrieved from https://www.cdc.gov/nchs/slaits/nsch.htm.

Thomas, R., Sanders, S., Doust, J., Beller, E., & Glasziou, P. (2015). Prevalence of Attention-Deficit/Hyperactivity Disorder: A Systematic Review and Meta-analysis. *Pediatrics,* peds.2014-3482. doi: 10.1542/peds.2014-3482
## 8. Paper Corpus
## 8.1 Included Papers

Antle, A. (2017) Crazy Like Us: Design for Vulnerable Populations. In *Proceedings of the 2017 Conference on Interaction Design and Children* (IDC '17), 3-4. doi:10.1145/3078072.3078074

Asiry, O., Shen, H., & Calder, P. (2015). Extending Attention Span of ADHD Children through an Eye Tracker Directed Adaptive User Interface. *In* Proceedings of the ASWEC 2015 24th Australasian Software Engineering Conference *(ASWEC ' 15 Vol. II)*, 149-152. doi: 10.1145/2811681.2824997

Beaton, R., Merkel, R., Prathipati, J., Weckstein, A., & McCrickard, S. (2014). Tracking mental engagement: a tool for young people with ADD and ADHD. In *Proceedings of the 16th international ACM SIGACCESS conference on Computers & accessibility* (ASSETS '14), 279-280. doi:10.1145/2661334.2661399

Dibia, V. (2016) FOQUS: A Smartwatch Application for Individuals with ADHD and Mental Health Challenges. In *Proceedings of the 18th International ACM SIGACCESS Conference on Computers and Accessibility* (ASSETS '16), 311-312. doi:10.1145/2982142.2982207

Duvall, J.C., Dunne, L.E., Schleif, N., & Holschuh, B. (2016). Active "hugging" vest for deep touch pressure therapy. In *Proceedings of the 2016 ACM International Joint Conference on Pervasive and Ubiquitous Computing: Adjunct* (UbiComp '16), 458-463. doi:10.1145/2968219.2971344

Eriksson, S., Gustafsson, F., Larsson, G., & Hansen, P. (2017). Takt: The Wearable Timepiece That Enables Sensory Perception of Time. In *Proceedings of the 2017 ACM Conference Companion Publication on Designing Interactive Systems* (DIS '17 Companion), 223-227. doi: 10.1145/3064857.3079150

Frutos-Pascual, M., Zapirain, B.G., & Buldian, K.C. (2014). Adaptive cognitive rehabilitation interventions based on serious games for children with ADHD using biofeedback techniques: assessment and evaluation. *In Proceedings of the 8th International Conference on Pervasive Computing Technologies for Healthcare (PervasiveHealth '14)*, 321-324. doi:10.4108/icst.pervasivehealth.2014.255249
39

## 8.2 Excluded Papers